\documentclass[titlepage,12pt]{article}
\usepackage{amssymb,psfig,epsfig,pslatex}
\def\shat{\hat{s}}

\def\sp{{\mathbf s}_t}
\newcommand{\sm}{{\mathbf s}_{\bar{t}}}
\newcommand{\kh}{{\hat{\mathbf k}}}
\newcommand{\ph}{\hat{\mathbf p}}

\newcommand{\one}{1\!\mbox{l}}
\begin{document}
\begin{titlepage}
\noindent
PITHA 01/02 \hfill April 2001 \\
DESY 01-040 \hfill \, \\
TTP01-11    \hfill \, 
\vspace{0.4cm}
\begin{center}
{\LARGE {\bf Next-to-leading order QCD corrections  
to \\ \vskip 0.2cm top quark spin
correlations
at hadron colliders: \\ \vskip 0.3cm the reactions 
\mbox{\boldmath $g  g \to t{\bar t}  (g)$}
and  \mbox{\boldmath $g q {(\bar q)} \to t{\bar t} q {(\bar q)}$}} } \\
\vspace{2cm}
{\bf W. Bernreuther $^{a}$,
A. Brandenburg $^{b,}$\footnote{supported by a Heisenberg fellowship of D.F.G.},
Z. G. Si $^{a,}$\footnote{supported by BMBF contract 05 HT9 PAA 1}
and P. Uwer $^c$}
\par\vspace{1cm}
$^a$ Institut f.\ Theoretische Physik, RWTH Aachen, 52056 Aachen, Germany\\
$^b$  DESY-Theorie, 22603 Hamburg, Germany\\
$^c$ Institut f. Theoretische Teilchenphysik, Universit\"at Karlsruhe,
76128 Karlsruhe, Germany
\par\vspace{3cm}
{\bf Abstract:}\\
\parbox[t]{\textwidth}
{We have computed the cross section for $t\bar t$ production
by gluon-gluon fusion  at next-to-leading order (NLO) in the QCD coupling,
keeping the full dependence on the $t\bar t$ spins.
Furthermore we have determined to the same order the
spin dependent cross sections for the processes 
$g + q ({\bar q})\to t {\bar t} + q
({\bar q})$.
Together with our previous results \cite{Bernreuther:2000yn}
for $q + {\bar q} \to t {\bar t} (g)$
these results allow for predictions, at NLO QCD, of the hadronic production
of  $t\bar t$ pairs in a general spin configuration.
As an application we have determined the degree of correlation of the
$t$ and $\bar t$ spins
 at NLO, using various spin quantisation axes. }
\end{center}
\vspace{2cm}
PACS number(s): 12.38.Bx, 13.88.+e, 14.65.Ha
\end{titlepage}
%
\noindent
In this Letter we report on the calculation of the cross section
for $t\bar t$ production by gluon-gluon and by  (anti)quark-gluon fusion
at order  $\alpha_s^3$ in the QCD coupling, keeping the full
information on the spin state of the $t\bar{t}$ system.
The results presented in this work constitute the last missing
ingredient
to analyse, at NLO QCD, $t\bar t$ production including spin effects at
hadron colliders.
Together with the corresponding results  for $t\bar t$ production by
quark-antiquark annihilation \cite{Bernreuther:2000yn}, they
extend
previous results for the spin-averaged differential $t\bar t$ cross
section
\cite{Nason:1988,Nason:1989,Beenakker:1989,Beenakker:1991}.

Analysis of top
quark spin phenomena at the upgraded Tevatron and at the Large Hadron
Collider (LHC)
will play an important role in investigating the interactions of these quarks.
(For a recent overview of the perspectives of top quark spin physics at
hadron colliders
and an extensive list of references,
see, e.g., ref.~\cite{Beneke:2000hk}.) In the following we first describe the
salient features of our
computation. Then we determine the correlation of the $t$ and $\bar t$ spins
at NLO for a number of spin bases.

The theoretical description of $t\bar t$ production by proton-proton and 
proton-antiproton collisions
at next-to-leading order in the QCD coupling amounts to considering the
following
parton reactions:  quark-antiquark annihilation
\begin{equation}
q(p_1) + {\bar q}(p_2) \rightarrow t(k_t) + {\bar t}(k_{\bar{t}}),
\label{eq:qq}
\end{equation}
\begin{equation}
q(p_1) + {\bar q}(p_2) \rightarrow t(k_t) + {\bar t}(k_{\bar{t}}) + g(k_3),
\label{eq:qqgluon}
\end{equation}
which is the dominant production mechanism at the Tevatron, gluon-gluon fusion
\begin{equation}
g(p_1) + g(p_2) \rightarrow t(k_t) + {\bar t}(k_{\bar{t}}),
\label{eq:gg}
\end{equation}
\begin{equation}
g(p_1) + g(p_2) \rightarrow t(k_t) + {\bar t}(k_{\bar{t}}) + g(k_3),
\label{eq:ggg}
\end{equation}
which dominates at the LHC, and
\begin{equation}
g(p_1) + q(p_2) \rightarrow t(k_t) + {\bar t}(k_{\bar{t}}) + q(k_3),
\label{eq:qg}
\end{equation}
\begin{equation}
g(p_1) + {\bar q}(p_2) \rightarrow t(k_t) + {\bar t}(k_{\bar{t}}) + {\bar q}(k_3).
\label{eq:qbarg}
\end{equation}

In the Standard Model the main top decay modes are
$t\to b W \to b q {\bar q}',  b \ell \nu_{\ell}$. Among these final states the
charged leptons, or the jets from quarks of weak isospin $-$1/2 originating
from the $W$ decay, are the most powerful analysers of
the polarisation of the top quark. A complete next-to-leading order  QCD
analysis thus consists of
treating the  parton reactions
\begin{equation}
gg, q{\bar q}  \rightarrow t{\bar t} \rightarrow b {\bar b}  + 4f,
\label{eq:ttrec1}
\end{equation}
\begin{equation}
gg, q{\bar q}  \rightarrow t{\bar t} \rightarrow b {\bar b}  + 4f + g,
\label{eq:ttrec2}
\end{equation}
\begin{equation}
g + q ({\bar q})  \rightarrow t{\bar t} \rightarrow b {\bar b}  + 4f + q
({\bar q}),
\label{eq:ttrec3}
\end{equation}
where $f=q,\ell,\nu_{\ell}$. To leading order in $\alpha_s$ only the reactions
(\ref{eq:ttrec1}) contribute.
In view of the fact that the total width
$\Gamma_t$
of the top quark is much smaller than its mass, $\Gamma_t/m_t ={\cal
O}(1\%)$, one may  analyse the above reactions
 using the so-called narrow width or leading pole approximation
\cite{Stuart:1991,Aeppli:1994}. In the  case at hand this approach
consists, for a given process,
 in an expansion of the amplitude around the poles of the unstable $t$ and
$\bar t$ quarks, which corresponds to
an expansion in powers of $\Gamma_t/m_t$. Only the leading term of this
expansion, i.e.,
the residue of the double poles is  considered here. In this framework the
radiative
corrections to (\ref{eq:ttrec1}) can be classified into so-called
factorisable and
non-factorisable
corrections. In the case of the gluon radiation
processes (\ref{eq:ttrec2}) already
the lowest order contributions to the respective squared matrix element
$\vert{\cal M}{\vert}^2$ decompose into these two groups.
 We compute the factorisable
 corrections\footnote{The non-factorisable NLO QCD contributions were
studied for
$g g$ and $q \bar q$ initial states in
ref.~\cite{Beenakker:1999}.}  for which
the squared
 matrix element ${\cal M}$ is
of the form
\begin{equation}
\vert{\cal M}{\vert}^2 \propto {\rm Tr}\;[\rho
R{\bar{\rho}}]
 = \rho_{\alpha'\alpha}
R_{\alpha\alpha',\beta\beta'}{\bar{\rho}}_{\beta'\beta} .
\label{eq:trace}
\end{equation}
Here $R$ denotes the  density matrix for the production of on-shell
$t\bar t$ pairs and
$\rho,{\bar{\rho}}$ are the density matrices describing the decay
of polarised $t$ and $\bar t$ quarks, respectively, into specific final states.
The subscripts in  eq.~(\ref{eq:trace}) denote the  $t$, $\bar t$ spin indices.
Because the
colours of the final state partons are summed over, $\rho,{\bar{\rho}}$ are
unit matrices
in the colour indices of  $t$ and $\bar t$. This implies that the
colour indices of  $t$ and $\bar t$ are summed over in the computation of $R$.
Note that both the production and decay density matrices are gauge invariant.
The one-loop QCD corrections to the  density matrices of semileptonic
decays of polarised top
quarks and of
$t \to W + b$ can be obtained from the results of refs.~\cite{Czarnecki:1991} 
and \cite{Schmidt:1996,Fischer:1999}, respectively.
\par
In the calculation of the production density matrices for the reactions
(\ref{eq:gg}) -- (\ref{eq:qbarg})
we follow our approach \cite{Bernreuther:2000yn} to computing those of
reaction (\ref{eq:qq}) and reaction (\ref{eq:qqgluon}).
The  density matrix for the production process (\ref{eq:gg})
is defined in terms of the transition matrix element as follows:
\begin{equation}
R^{(gg)}_{\alpha\alpha' ,\beta\beta'}=
\frac{1}{N_{gg}}
\sum_{{{\rm\scriptscriptstyle colors} \atop
{\rm\scriptscriptstyle initial}\;
{\rm\scriptscriptstyle spins} }}
\langle t_\alpha\bar t_\beta |{\cal T}|
\,g g\,\rangle\;
\langle \,g g\,|{\cal T}^\dagger|
t_{\alpha'}\bar t_{\beta'}\rangle\;,
\label{eq:Rdef}
\end{equation}
where
the factor $N_{gg} = 256$
averages over the spins and colours
of the initial pair of gluons.
The matrix structure of $R^{(gg)}$  is
\par
\hfill\parbox{13.4cm}{
\begin{eqnarray*}
R^{(gg)}_{\alpha\alpha',\beta\beta'}&=&
A^{(gg)} \delta_{\alpha\alpha'}\delta_{\beta\beta'}
+B^{(gg)}_{i} (\sigma^i)_{\alpha\alpha'}
\delta_{\beta\beta'} +{\bar B}^{(gg)}_{i} \delta_{\alpha\alpha'}
(\sigma^i)_{\beta\beta'} \\
&&+\, C^{(gg)}_{ij}(\sigma^i)_{\alpha\alpha'}
(\sigma^j)_{\beta\beta'}  \,\, ,
\label{eq:Rstruct}
\end{eqnarray*} }\hfill\parbox{0.8cm}{\begin{eqnarray}  \end{eqnarray} }
\par\noindent
where $\sigma^i$ are the Pauli matrices. Using rotational invariance the
`structure functions'  $B^{(gg)}_i,{\bar B}^{(gg)}_i$ and
$C^{(gg)}_{ij}$ can be
further decomposed. The function
$A^{(gg)}={\rm Tr}\;[R^{(gg)}]/4$,  
determines the $gg \to t{\bar t}$ differential cross section
with $t{\bar t}$ spins summed over. 
Because of parity  invariance  the vectors  
${\bf B}^{(gg)},{\bf\bar B}^{(gg)}$ can
have, within QCD, 
only a component normal to the scattering plane. This component,
which amounts to a normal polarisation of the $t$  and $\bar t$ quarks,
is induced by the absorptive part of the  scattering amplitude which
is non-zero  to order $\alpha^3_s$.
The normal polarisation is quite small, both for
$t\bar t$ production
at the Tevatron and at the LHC \cite{Bernreuther:1996,Dharmaratna:1996}.
The functions $C^{(gg)}_{ij}$ encode the correlation  between the  $t$ and
${\bar t}$ spins.
The production density matrices $R^{(gg,g)}$, $R^{(gq,q)}$, and $R^{(g{\bar
q},{\bar q})}$
for the reactions
 (\ref{eq:ggg}),  (\ref{eq:qg}), and (\ref{eq:qbarg}),
 can be defined and decomposed in an analogous fashion,
with the colour and spin degrees of freedom of the final state
gluon, quark $q$, or antiquark $\bar{q}$ being summed over.

In Born approximation  $R^{(gg)}$ was given, e.g., in
ref.~\cite{Brandenburg:1996}.
In the computation of the
order $\alpha_s^3$ contributions  we used
dimensional regularization to treat both the
ultraviolet
and the infrared/collinear singularities.
 Renormalisation was performed using
 the $\overline{\rm{MS}}$ prescription for the QCD coupling $\alpha_s$
and the on-shell definition of the top mass $m_t$.
The remaining soft and collinear singularities in  $R^{(gg)}$, which
appear as single and double poles in $\epsilon = (4-D)/2$,
 are cancelled after including the contributions of $R^{(gg,g)}$
in the soft and collinear limits
 and  after mass factorisation. For the latter we
used the $\overline{\rm{MS}}$ factorisation scheme.
The soft and collinear singularities of $R^{(gg,g)}$
were extracted by employing
a simplified version of the phase-space slicing technique (cf.
ref.~\cite{Bernreuther:2000yn}
for details).
\par
As a check of our calculation we computed the total cross section for
$g g \to t{\bar t} + X$ at NLO.
If one identifies the $\overline{\rm{MS}}$ renormalisation scale $\mu$
with the mass factorisation scale $\mu_F$  and neglects all quark
masses except for $m_t$, then one can express the parton cross section in
 terms of dimensionless scaling functions \cite{Nason:1988}:
\begin{equation}
\hat{\sigma}_{gg}(\shat,m^2_t)=\frac{\alpha_s^2}{m^2_t}[
f^{(0)}_{gg}(\eta) + 4\pi\alpha_s(f^{(1)}_{gg}(\eta) +
{\tilde f}^{(1)}_{gg}(\eta) \ln(\mu^2/m^2_t))],
\label{eq:xsection}
\end{equation}
where $\shat$ is the parton c.m. energy squared and 
\begin{equation}
\eta = \frac{\shat}{4m^2_t} -1.
\end{equation}
Our results for these functions are shown in Fig.~1.
The dotted line is the Born result $f^{(0)}_{gg}(\eta)$, the full
line shows the function $f^{(1)}_{gg}(\eta)$, and the dashed
line is ${\tilde f}^{(1)}_{gg}(\eta)$.
\begin{figure}
\unitlength1.0cm
\begin{center}
\begin{picture}(12,12)
\put(0,0){\psfig{figure=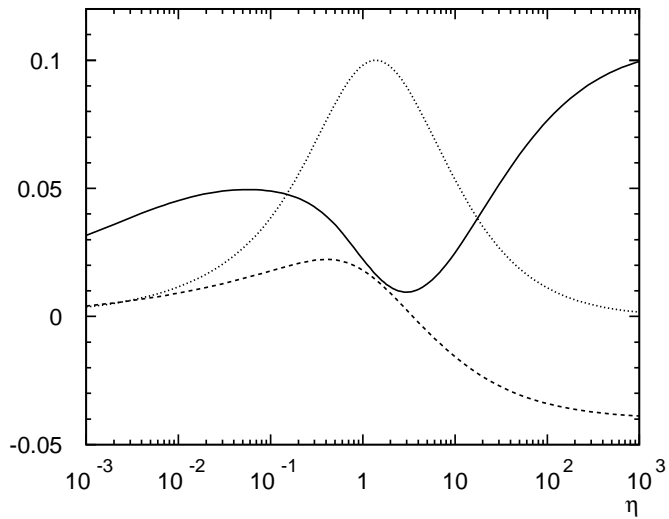,width=9cm,height=9cm}}
\end{picture}
\vskip -3.5cm
\caption{Dimensionless scaling functions $f^{(0)}_{gg}(\eta)$
(dotted), $f^{(1)}_{gg}(\eta)$ (full), and
${\tilde f}^{(1)}_{gg}(\eta)$ (dashed) that determine
parton cross section $\hat{\sigma}_{gg}$.}\label{fig:siggg}
\end{center}
\end{figure}
\begin{figure}
\unitlength1.0cm
\begin{center}
\begin{picture}(12,12)
\put(0,0){\psfig{figure=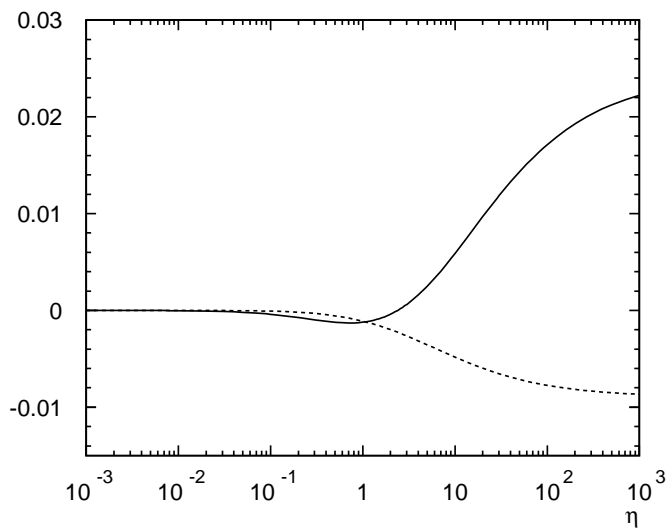,width=9cm,height=9cm}}
\end{picture}
\vskip -3.5cm
\caption{Dimensionless scaling functions  
$f^{(1)}_{gq}(\eta)$ (full)  and
${\tilde f}^{(1)}_{gq}(\eta)$ (dashed) that determine $\hat{\sigma}_{gq}$.}\label{fig:sigqg}
\end{center}
\end{figure}
We find excellent agreement
with the corresponding results for $f^{(0,1)}_{gg}(\eta)$
of ref.~\cite{Nason:1988,Beenakker:1989} and, after accounting for the
different renormalisation schemes used by us and in ref.~\cite{Nason:1988}, 
with ${\tilde f}^{(1)}_{gg}(\eta)$.
\par
The determination of $R^{(gq,q)}$ to order
$\alpha_s^3$,
describing the (anti)quark-gluon subprocesses (\ref{eq:qg}) and
(\ref{eq:qbarg}),
involves initial state collinear singularities that we
mass-factorised using the $\overline{\rm{MS}}$ factorisation scheme.
The  parton cross section  $\hat{\sigma}_{gq}(\shat,m^2_t)$
takes the form
\begin{equation}
{\hat\sigma_{gq}}  = \frac{4 \pi\alpha_s^3}{m_t^2}
 [ f^{(1)}_{gq}(\eta) +
{\tilde f}^{(1)}_{gq}(\eta) \ln(\mu^2/m^2_t))] \, .
\label{eq:siggq}
\end{equation}
The functions $f^{(1)}_{gq}(\eta)$ (full line) 
and ${\tilde f}^{(1)}_{gq}(\eta)$ (dashed line) 
are shown in Fig.~2. 
They agree
perfectly with the corresponding results of
ref.~\cite{Nason:1988,Beenakker:1991}. 
The density matrix $R^{(g\bar{q},\bar{q})}$ can be obtained
from  $R^{(gq,q)}$ 
by exploiting the charge-conjugation invariance of QCD.
\par
In order to exhibit the $t\bar t$ spin-correlation effects
we consider now the following set of
observables:
\begin{equation}
{\cal O}_1=4\,\sp\cdot\sm,
\label{eq:sbasis}
\end{equation}
\begin{equation}
{\cal O}_2=4\,(\kh_t\cdot\sp)(\kh_{\bar{t}}\cdot\sm),
\label{eq:hbasis}
\end{equation}
\begin{equation}
{\cal O}_3=4\,(\ph_1\cdot\sp)(\ph_1\cdot\sm),
\label{eq:pbasis}
\end{equation}
\begin{equation}
{\cal O}_4=4\,(\ph_2^*\cdot\sp)(\ph_1^{**}\cdot\sm),
\label{eq:ybasis}
\end{equation}
where $\sp=(\sigma\otimes \one)/2$ 
and $\sm=(\one\otimes\sigma)/2$ are the $t$ and $\bar{t}$ spin operators
and the factor 4 is conventional.
The expectation values of the  observables
${\cal O}_2$, ${\cal O}_3$ and ${\cal O}_4$ 
 determine the $t\bar{t}$
spin correlations using different spin quantisation axes.
We consider here the fully inclusive case, i.e. 
integrate over the appropriate full final state phase space.
Observable ${\cal O}_2$ corresponds to a
correlation of the $t$ and $\bar t$ spins
in the helicity basis, while  ${\cal O}_3$ correlates the spins projected
along the beam line in the parton c.m.s.
The `beam-line basis'  \cite{Mahlon:1996} used in  ${\cal O}_4$
 refers to spin axes being parallel to
  $\ph_2^*$ and $\ph_1^{**}$, respectively.
Here $\ph_2^*$ ($\ph_1^{**}$) is the direction of the downward
(upward) collider beam
in the $t$ ($\bar t$) rest frame which we define 
by performing a rotation-free boost from the parton c. m. frame.
The expectation values of ${\cal O}_2$, ${\cal O}_3$, and ${\cal O}_4$ are
related to the perhaps more familiar double spin asymmetries
of the cross section as follows:
\begin{equation}
  4 \langle ({\bf{a}}\cdot \sp)({\bf{b}}\cdot \sm) \rangle
  = {\sigma(\uparrow \uparrow)+\sigma(\downarrow \downarrow)
  - \sigma(\uparrow \downarrow)- \sigma(\downarrow \uparrow)\over
  \sigma(\uparrow \uparrow)+\sigma(\downarrow \downarrow)
  + \sigma(\uparrow \downarrow)+ \sigma(\downarrow \uparrow)
  },
\end{equation}
where the arrows refer to the spin state of the top quark (antiquark)
using the unit vector ${\bf{a}}$ (${\bf{b}}$) as quantisation axis.
\par
For the $gg$ initial state we write the unnormalised expectation value of a
spin-correlation
observable ${\cal O}$
 in analogy to eq.~(\ref{eq:xsection})
as follows:
\begin{equation}
{\hat\sigma_{gg}} \langle {\cal O} \rangle_{gg}  = \frac{\alpha_s^2}{m_t^2}
[ g^{(0)}_{gg}(\eta) + 4\pi\alpha_s(g^{(1)}_{gg}(\eta) +
{\tilde g}^{(1)}_{gg}(\eta) \ln(\mu^2/m^2_t))] \, ,
\label{eq:expval}
\end{equation}
\begin{figure}
\unitlength1.0cm
\begin{center}
\begin{picture}(12,12)
\put(0,0){\psfig{figure=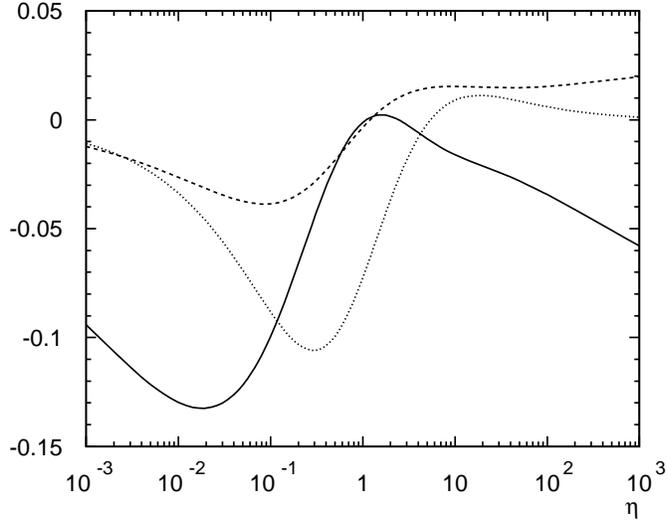,width=9cm,height=9cm}}
\end{picture}
\vskip -3.5cm
\caption{Dimensionless scaling functions $g^{(0)}_{gg}(\eta)$
(dotted), $g^{(1)}_{gg}(\eta)$ (full), and
${\tilde g}^{(1)}_{gg}(\eta)$ (dashed) that determine
the expectation value $\hat{\sigma}_{gg}\langle {\cal O}_1 \rangle_{gg}$.}\label{fig:o1}
\end{center}
\end{figure}
\begin{figure}
\unitlength1.0cm
\begin{center}
\begin{picture}(12,12)
\put(0,0){\psfig{figure=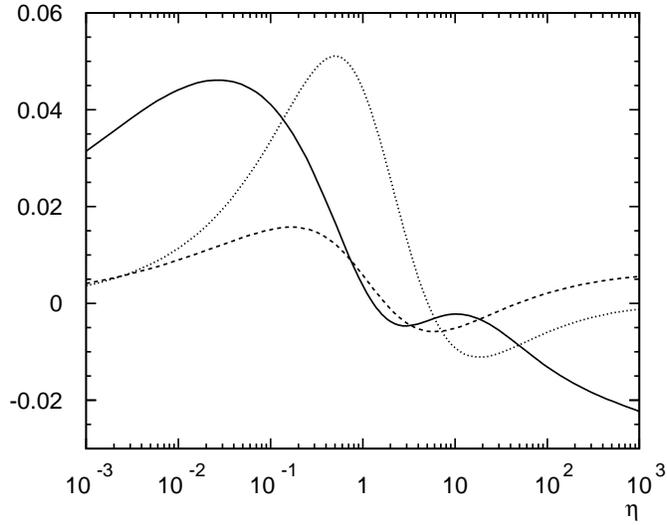,width=9cm,height=9cm}}
\end{picture}
\vskip -3.5cm
\caption{Same as Fig.1, but for $\hat{\sigma}_{gg}\langle {\cal O}_2
\rangle_{gg}$.}\label{fig:o2}
\end{center}
\end{figure}
\begin{figure}
\unitlength1.0cm
\begin{center}
\begin{picture}(12,12)
\put(0,0){\psfig{figure=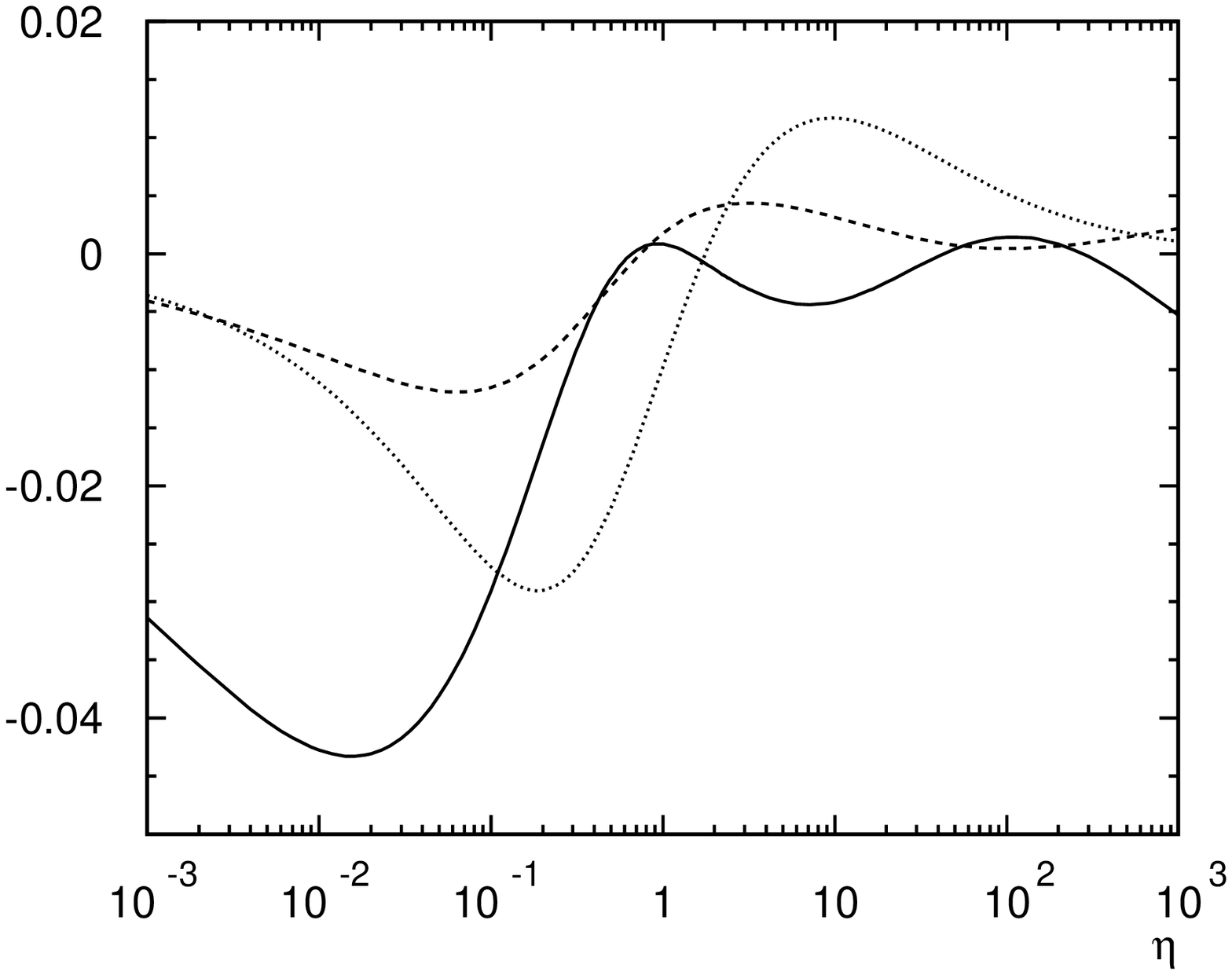,width=9cm,height=9cm}}
\end{picture}
\vskip -3.5cm
\caption{Same as Fig.1, but for $\hat{\sigma}_{gg}\langle {\cal O}_3
\rangle_{gg}$.}\label{fig:o3}
\end{center}
\end{figure}
\begin{figure}
\unitlength1.0cm
\begin{center}
\begin{picture}(12,12)
\put(0,0){\psfig{figure=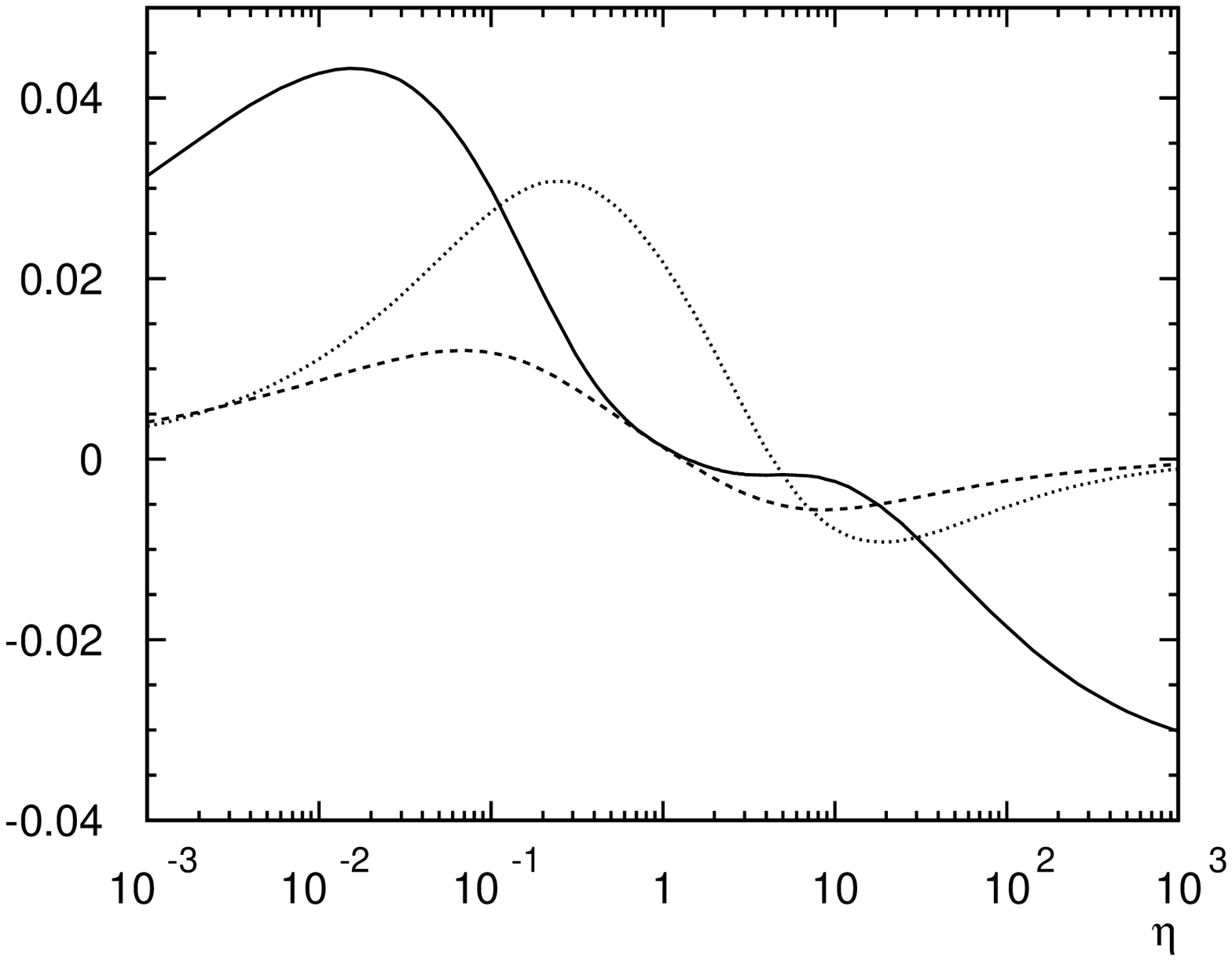,width=9cm,height=9cm}}
\end{picture}
\vskip -3.5cm
\caption{Same as Fig.1, but for $\hat{\sigma}_{gg}\langle {\cal O}_4
\rangle_{gg}$.}\label{fig:o4}
\end{center}
\end{figure}
\begin{figure}
\unitlength1.0cm
\begin{center}
\begin{picture}(12,12)
\put(0,0){\psfig{figure=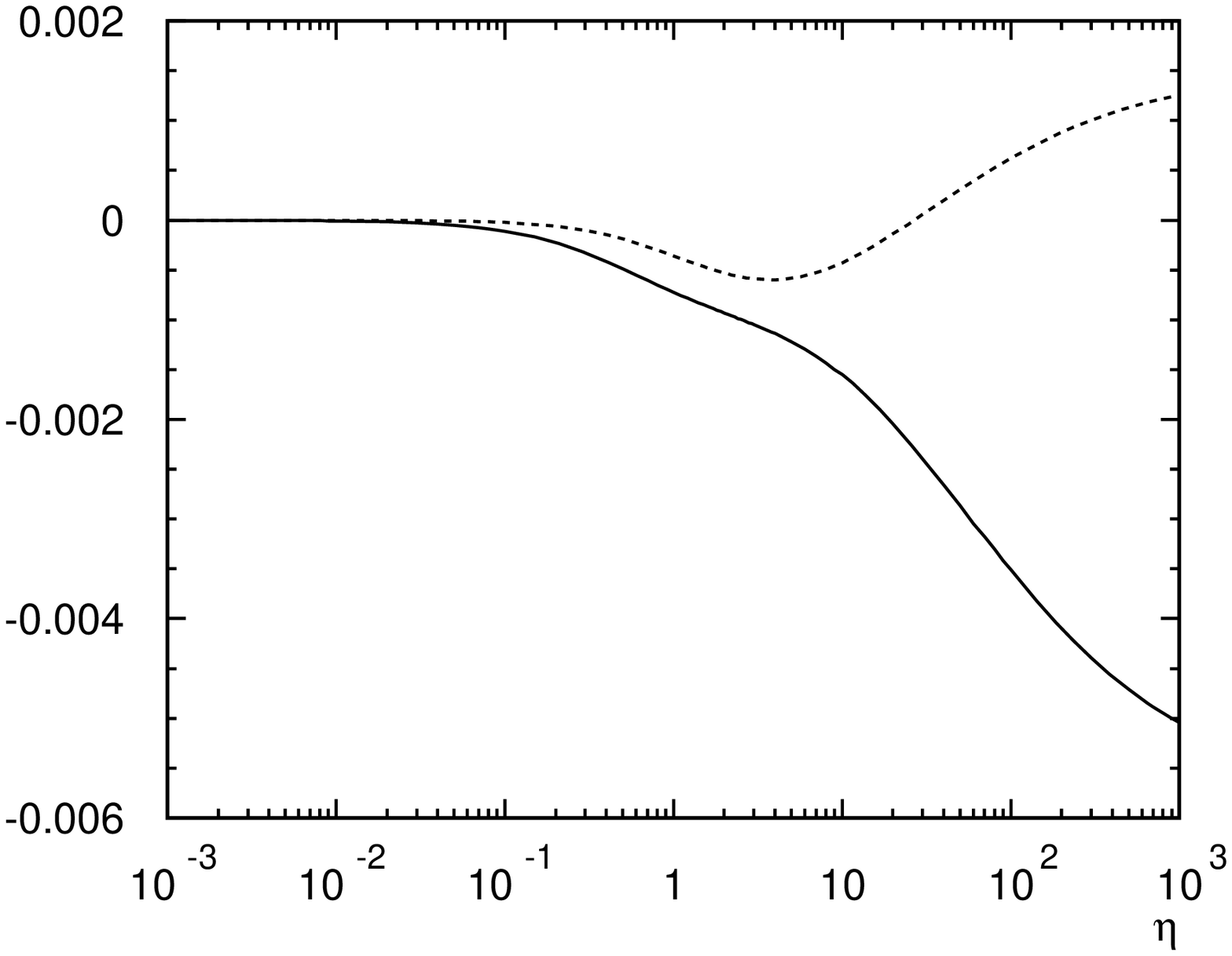,width=9cm,height=9cm}}
\end{picture}
\vskip -3.5cm
\caption{Dimensionless scaling functions $g^{(1)}_{gq}(\eta)$
(full) and
${\tilde g}^{(1)}_{gq}(\eta)$ (dashed) that determine
the expectation value $\hat{\sigma}_{gq}\langle {\cal O}_2 \rangle_{gq}$.}\label{fig:o2_gq}
\end{center}
\end{figure}
\begin{figure}
\unitlength1.0cm
\begin{center}
\begin{picture}(12,12)
\put(0,0){\psfig{figure=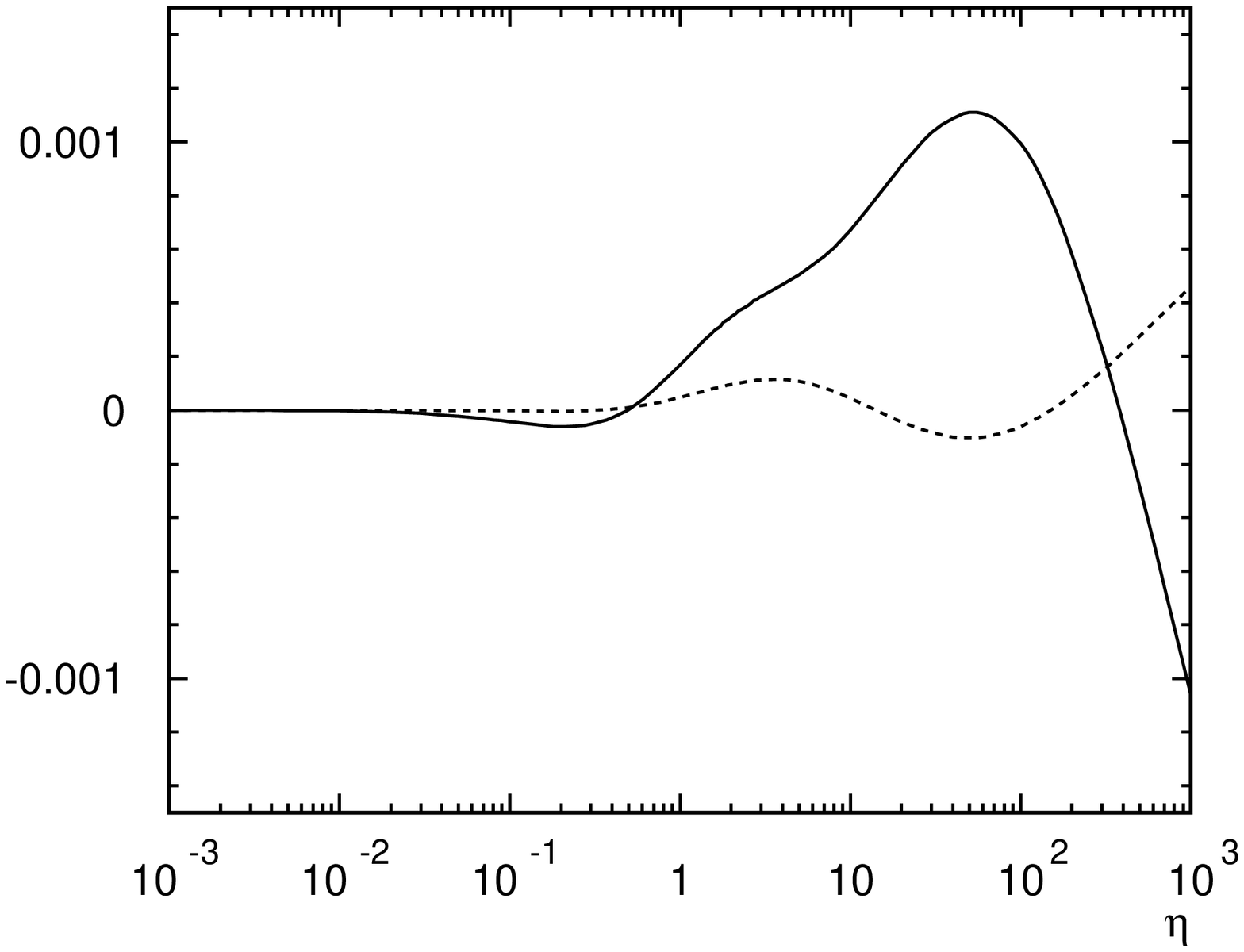,width=9cm,height=9cm}}
\end{picture}
\vskip -3.5cm
\caption{Same as Fig.7, but for $\hat{\sigma}_{gq}\langle {\cal O}_3
\rangle_{gq}$.}\label{fig:o3_gq}
\end{center}
\end{figure}
Our results for the functions
$g^{(0)}_{gg}(\eta),\ g^{(1)}_{gg}(\eta)$ and
${\tilde g}^{(1)}_{gg}(\eta)$ are shown for the four observables
${\cal O}_1$--${\cal O}_4$ in Figs. 3--6. In each figure,
the dotted line is the Born result $g^{(0)}_{gg}(\eta)$, the full
line shows the function $g^{(1)}_{gg}(\eta)$, and the dashed
line is ${\tilde g}^{(1)}_{gg}(\eta)$.
\par
The QCD corrections to the unnormalised spin correlations are large close
to threshold. This behaviour is due to the factor $\hat{\sigma}_{gg}$, 
which, in this order of perturbation theory, is non-zero at
threshold due to Coulomb attraction. 
The corrections to the normalised partonic expectation values of all spin 
observables are tiny for small $\eta$. This is also the case for the
contributions from $q\bar{q}$ annihilation \cite{Bernreuther:2000yn}.
For larger $\eta$, the spin correlations show a rich structure and the 
QCD corrections to the normalised expectation values can become significant.
For hadronic observables, these effects are damped by the parton distribution
functions and the QCD corrections are of the order of a few percent
\cite{Bernreuther:2001}.
\par
For the quark-gluon process (\ref{eq:qg}) we represent the
unnormalised expectation value of an observable in the form
\begin{equation}
{\hat\sigma_{gq}} \langle {\cal O} \rangle_{gq}  = \frac{4 \pi\alpha_s^3}{m_t^2}
 [ g^{(1)}_{gq}(\eta) +
{\tilde g}^{(1)}_{gq}(\eta) \ln(\mu^2/m^2_t))] \, .
\label{eq:expgq}
\end{equation}

We have plotted in Figs. 7,8 the functions  $g^{(1)}_{gq}(\eta)$, 
${\tilde g}^{(1)}_{gq}(\eta)$
corresponding to the two observables ${\cal O}_2$, ${\cal O}_3$.
The contributions of this channel to the spin correlations are very small.

\par
In summary, we have computed the
spin density matrices describing $t\bar{t}$ production
by gluon-gluon and by (anti)quark-gluon fusion to order $\alpha_s^3$.
Together with our previous result on $R^{q\bar q}$ at NLO
this work now provides the ingredients for a complete description 
of spin effects in the hadronic production of top quark pairs at NLO in
the strong coupling, allowing for a more precise investigation of the 
$t\bar{t}$ production and decay processes.


%

\begin{thebibliography}{99}


\bibitem{Bernreuther:2000yn}
W.~Bernreuther, A.~Brandenburg and Z.~G.~Si,
Phys.\ Lett.\ B483 (2000) 99
[hep-ph/0004184].



\bibitem{Nason:1988}
P.~Nason, S.~Dawson and R.~K.~Ellis,
Nucl.\ Phys.\  B\ 303 (1988) 607.

\bibitem{Nason:1989}
P.~Nason, S.~Dawson and R.~K.~Ellis,
Nucl.\ Phys.\  B\ 327  (1989) 49.

\bibitem{Beenakker:1989}
W.~Beenakker, H.~Kuijf, W.~L.~van Neerven and J.~Smith,
Phys.\ Rev.\  D\ 40 (1989) 54.

\bibitem{Beenakker:1991}
W.~Beenakker, W.~L.~van Neerven, R. Meng, G. A. Schuler and J.~Smith,
Nucl.\ Phys.\ B\ 351 (1991) 507.

\bibitem{Beneke:2000hk}
M.~Beneke {\it et al.},
hep-ph/0003033.


\bibitem{Stuart:1991}
R. G. Stuart, Phys. Lett. B 262 (1991) 113.

\bibitem{Aeppli:1994}
A. Aeppli, G. J. van Oldenborgh and D. Wyler, Nucl. Phys. B 428 (1994) 126
[hep-ph/9312212].

\bibitem{Beenakker:1999}
W.~Beenakker, F. A. Berends and A. P. Chapovsky,
Phys.\ Lett.\ B\ 454 (1999) 129 [hep-ph/9902304].

\bibitem{Czarnecki:1991}
A.~Czarnecki, M.~Jezabek and J.~H.~K\"uhn,
Nucl.\ Phys.\ B 351 (1991) 70.

\bibitem{Schmidt:1996}
C.~R.~Schmidt,
Phys.\ Rev.\   D 54 (1996) 3250 [hep-ph/9504434].

\bibitem{Fischer:1999}
M.~Fischer, S.~Groote, J.~G.~K\"orner, M.~C.~Mauser and B.~Lampe,
Phys.\ Lett.\   B 451 (1999) 406 [hep-ph/9811482].

\bibitem{Bernreuther:1996}
W.~Bernreuther, A.~Brandenburg and P.~Uwer,
Phys.\ Lett.\   B\ 368 (1996) 153 [hep-ph/9510300].

\bibitem{Dharmaratna:1996}
W.~G.~Dharmaratna and G.~R.~Goldstein,
Phys.\ Rev.\  D\ 53 (1996) 1073.

\bibitem{Brandenburg:1996}
A.~Brandenburg,
Phys.\ Lett.\   B\ 388 (1996) 626 [hep-ph/9603333].

\bibitem{Mahlon:1996}
G.~Mahlon and S.~Parke,
Phys.\ Rev.\   D\ 53 (1996) 4886 [hep-ph/9512264] .

\bibitem{Bernreuther:2001}
W.~Bernreuther, A.~Brandenburg, Z.~G.~Si, and P.~Uwer,
work in progress.

\end{thebibliography}
\end{document}